# Multi-orbital charge density wave excitations and concomitant phonon anomalies in $Bi_2Sr_2LaCuO_{6+\delta}$

Jiemin Li[1,2], Abhishek Nag[1], Jonathan Pelliciari[3,4], Hannah Robarts[1,5], Andrew Walters[1], Mirian Garcia-Fernandez[1], Hiroshi Eisaki[6], Dongjoon Song[6], Hong Ding[2], Steven Johnston[7], Riccardo Comin[3], Ke-Jin Zhou[1]*

[1]Diamond Light Source, Harwell Campus, Didcot OX11 0DE, United Kingdom

[2]Beijing National Laboratory for Condensed Matter Physics and Institute of Physics, Chinese Academy of Sciences, Beijing 100190, China

[3]Department of Physics, Massachusetts Institute of Technology, Cambridge, Massachusetts 02139, USA

[4]National Synchrotron Light Source II, Brookhaven National Laboratory, Upton, NY 11973, USA

[5]H. H. Wills Physics Laboratory, University of Bristol, Bristol BS8 1TL, United Kingdom

[6]National Institute of Advanced Industrial Science and Technology (AIST), Tsukuba, Ibaraki 305-8560, Japan

[7]Department of Physics and Astronomy, The University of Tennessee, Knoxville, TN 37996, USA

**Correspondence to: kejin.zhou@diamond.ac.uk**

Charge density waves (CDWs) are ubiquitous in under-doped cuprate superconductors. As a modulation of the valence electron density, CDWs in hole-doped cuprates possess both Cu-3$d$ and O-2$p$ orbital character owing to the strong hybridization of these orbitals near the Fermi level. Here, we investigate under-doped Bi$_2$Sr$_{1.4}$La$_{0.6}$CuO$_{6+\delta}$ using resonant inelastic X-ray scattering (RIXS) and find that a short-range CDW exists at both Cu and O sublattices in the copper-oxide (CuO$_2$) planes with a comparable periodicity and correlation length. Furthermore, we uncover bond-stretching and bond-buckling phonon anomalies concomitant to the CDWs. Comparing to slightly over-doped Bi$_2$Sr$_{1.8}$La$_{0.2}$CuO$_{6+\delta}$, where neither CDWs nor phonon anomalies appear, we highlight that a sharp intensity anomaly is induced in the proximity of the CDW wavevector (Q$_{CDW}$) for the bond-buckling phonon, in concert with the diffused intensity enhancement of the bond-stretching phonon at wavevectors much greater than Q$_{CDW}$. Our results provide a comprehensive picture of the quasi-static CDWs, their dispersive excitations, and associated electron-phonon anomalies, which are key for understanding the competing electronic instabilities in cuprates.

**Introduction**

Cuprate superconductivity is achieved through doping into the stoichiometric parent compound, which is a charge-transfer Mott insulator with strong electronic correlations. Here, the injected holes can segregate into periodically spaced domain walls, rapidly suppressing static antiferromagnetic (AFM) order. A classic example is the ordered stripe phase in $La_{1.48}Nd_{0.4}Sr_{0.12}CuO_4$ (LNSCO), where both spin and charge form a quasi-static order near a doping level of ~1/8 (1). Recently, quasi-two-dimensional short-range charge density waves (CDWs) have been discovered in most hole-doped cuprates (2-12). Though hole carriers are primarily doped into the O-2$p$ orbitals in the $CuO_2$ planes, resonant elastic X-ray scattering (REXS) experiments have detected the quasi-static CDW order within both the Cu and O sublattices in $La_{1.875}Ba_{0.125}CuO_4$ (LBCO) owing to the strong hybridization between the Cu-3$d$ and the O-2$p$ orbitals (2, 13-14). The CDW in LBCO is, however, relatively long-range, leaving an open question on whether generic short-range CDWs (such as the Bi-based compounds) also project onto multiple orbitals.

Subsequent to the discovery of the stripe phase in LNSCO, inelastic neutron scattering (INS) identified a pronounced softening in the Cu-O bond-stretching phonon branch close to the CDW wavevector in La-based cuprates and $YBa_2Cu_3O_{6+\delta}$ (YBCO) (15). In $Bi_2SrLaCuO_{6+\delta}$, non-resonant inelastic X-ray scattering (IXS) studies also revealed softening in the bond-stretching phonon (16). Most recently, a similar electron-phonon anomaly was uncovered by RIXS in under-doped $Bi_2Sr_2CaCu_2O_{8+\delta}$ (Bi2212), where a short-range CDW occurs (17). These studies suggest an intimate link between electron-phonon coupling (EPC) and CDW correlations. Interestingly, a dynamical dispersive CDW excitation is inferred in Bi2212 as an anomalously enhanced phonon intensity upon the interference between phonons and underlying charge excitations (17). It is unclear, however, whether such dispersive CDW excitations are ubiquitous to other cuprate families.

To date, the underlying mechanism of CDWs in cuprates still remains elusive. Unveiling the nature of the quasi-static CDWs as well as their dynamical excitations is crucial for understanding the CDW's origin – similar to the study of AFM in cuprates. Though the quasi-static properties of CDWs have been investigated extensively, their dynamics (*i.e.* the collective CDW excitations), are largely unexplored due to the limited availability of suitable experimental tools. RIXS is one of the few techniques that directly probe quasi-static CDWs, their excitations, and electron-phonon coupling, which are important elements for identifying the CDW mechanism.

**Results and discussions**

We initially used RIXS to study an under-doped superconducting single-layer $Bi_2Sr_{1.4}La_{0.6}CuO_{6+\delta}$ (Bi2201) (superconducting temperature $T_c$ = 23 K, UD23) (Methods and *SI Appendix,* Fig. S1). Figures 1A and 1B show the Cu $L_3$ and O $K$ X-ray absorption spectra (XAS) of UD23 with the incoming linear polarization (σ) parallel to the $CuO_2$ planes (σ polarized light is used throughout unless otherwise stated) (see *SI Appendix*, Fig. S2 for the experimental geometry). The Cu $L_3$ and O $K$ XAS represents a projection of the unoccupied states of planar Cu 3$d$ and O 2$p$ orbitals, respectively. The momentum dependence of the Cu $L_3$ RIXS measurements was acquired at the Cu $L_3$ resonance ($E \sim 931.6$ eV) and over a broad range of in-plane momentum transfers $q_{//}$. Figure 1C presents the low energy excitations (< 100 meV) in the Cu $L_3$ RIXS spectra as a function of $q_{//}$ = (H, 0). A quasi-elastic scattering peak is clearly visible at $q_{//} \sim 0.25$ reciprocal lattice units (*r.l.u.*). Such a peak was also seen in RIXS spectra under different configurations (see *SI Appendix*, Fig. S3). O $K$ RIXS spectra were collected with incident photon energy tuned to the resonance of the mobile carrier (hole) peak ($E \sim 528.4$ eV) as a function of $q_{//}$ = (H, 0) and up to ~ 0.3 *r.l.u.*. Figure 1D shows the corresponding data, where a quasi-elastic scattering peak also appears at $q_{//} \sim 0.25$ *r.l.u.*.

To clarify the origin of these peaks, we performed an energy-dependent RIXS scan at a fixed $q_{//}$ at both resonances (Cu $L_3$ and O $K$). The integrated intensities of the quasi-elastic peak are superimposed on top of the XAS data in Figs. 1A-1B. The resonance enhancement of the intensities at the Cu $L_3$ and O $K$ hole peaks unambiguously proves the existence of a CDW on both the Cu and O sublattices of the $CuO_2$ planes (2, 6, 11, 14). Notably, an energy offset exists between the maximum of the XAS spectra and the peak of the energy-dependent scattering intensities of the quasi-static CDW. A comparable offset has been seen in multiple cuprate CDW studies where the XAS connects with the imaginary part of the atomic form factor whilst the scattering relates to the real part of the form factor (2, 14, 18). For a more quantitative assessment, we plot the integrated intensity of the quasi-elastic peak in Fig. 1E. The peak position of the CDW is extracted to be $\sim 0.259 \pm 0.006$ *r.l.u.* and $\sim 0.25 \pm 0.003$ *r.l.u.* at Cu $L_3$ and O $K$ edges, respectively. The full width at the half-maximum (FWHM) of the CDW peak is $\Gamma \sim 0.078 \pm 0.006$ *r.l.u.* at the Cu $L_3$- and $\Gamma \sim 0.076 \pm 0.006$ *r.l.u.* at O $K$- edges. The corresponding CDW correlation length is $2/\Gamma \sim 15$ Å. Note that a comparable difference of the CDW wavevector between the Cu and O sublattices was also observed in REXS results from LBCO and is due to a tiny change in the refractive index of X-rays between the O $K$ and the Cu $L_3$ edges (2). We confirmed these observations on a second sample of UD23 and obtained comparable results (see *SI Appendix*, Fig. S4). The in-plane CDW wavevector and the FWHM of the CDW peak at the Cu $L_3$ are consistent with data obtained on similar Bi2201 compounds (6, 11). We conclude that the observed CDWs at the Cu and O sublattices have comparable periodicity and correlation length, reflecting the projection of the same electronic order onto multiple orbitals.

In Fig. 1C, we highlight an excitation at ~60 meV whose intensity is enhanced for $q_{//}$ between $Q_{CDW}$ and the zone boundary. At the O $K$ (Fig. 1D) the excitation is much weaker in comparison to the CDW peak. To quantify the $q_{//}$ dependence of the inelastic component, we fitted the quasi-elastic peak and then subtracted it from the RIXS spectra (Method, and *SI Appendix,* Figs. S5-S6). Fitting examples of the excitation spectra at the Cu $L_3$- and O $K$-edges are shown in Fig. 2A and Fig. 2B, respectively. The $q_{//}$-dependent inelastic excitations are displayed in Figs. 2C-2D with selected spectra depicted in Figs. 2E-2F covering $q_{//}$ of 0.2 *r.l.u.* to 0.3 *r.l.u.*. Most notably, the line profile of the inelastic spectra at the Cu $L_3$-edge is single-peaked while a double-peaked structure appears at the O $K$-edge. To elucidate the latter, we collected higher energy resolution RIXS spectra from the same sample and were able to clearly resolve the lower-energy peak (*SI Appendix,* Fig. S7). The dispersion and integrated spectral weight were extracted and summarized in Figs. 2G-2J. At the Cu $L_3$ resonance (Fig. 2G), the inelastic peak is located around 65 meV at small $q_{//}$ then softens in the $q_{//}$ range of 0.2 ~ 0.4 *r.l.u.*, with a broad 'dip' developing at about 50 meV near $Q_{CDW}$. The extracted dispersion matches well with those obtained using π polarized light and they are reminiscent of the phonon softening observed in Bi2212 (17) (*SI Appendix,* Fig. S8). Although the softening wavevector is close to $Q_{CDW}$ in Bi2201 and Bi2212, a recent RIXS study of CDW correlations in LSCO shows that the phonon softening develops at Q > $Q_{CDW}$, possibly implying a delicate relationship between the momentum of the phonon softening and the CDW wavevectors (19). In Fig. 2H, the integrated intensity has a non-monotonic increase as a function of $q_{//}$ with a maximum around 0.35 *r.l.u.*. Noticeably, both the dispersion and the intensity profiles of the inelastic peak agree very well with that of the Cu-O bond-stretching phonon branch (the half-breathing $\Delta_1$ mode along the (100) direction) in the under-doped Bi2212 (17). The intensity enhancement at wavevectors larger than $Q_{CDW}$ resembles the Fano interference effect of the bond-stretching phonon in the under-doped Bi2212 (17). These observations suggest the existence of the dispersive CDW excitations that interact with the bond-stretching phonon in UD23. We highlight that the phonon intensities measured by RIXS and IXS are very different: the former is proportional to the strength of the electron-phonon coupling (EPC) in the reciprocal space while the latter measures the phonon self-energy (15-17, 20-21).

Concerning the excitation at the O $K$-edge, the high-energy peak exhibits a downward dispersion (Fig. 2I) and a rising intensity (Fig. 2J) with increasing $q_{//}$, akin to the bond-stretching phonon branch observed at the Cu $L_3$-edge. The low-energy peak, however, is centered at ~ 30 meV showing little dispersion (Fig. 2I). Its intensity (Fig. 2J) gradually decreases from $q_{//}$ of 0.1 $r.l.u.$ to 0.2 $r.l.u.$, before abruptly forming a dome-shaped enhancement peaked around $Q_{CDW}$ of 0.25 $r.l.u.$. The central energy of the low-energy branch is comparable to that of a bond-buckling phonon in YBa$_2$Cu$_3$O$_7$ (YBCO) identified by INS (22). In fact, a RIXS study on the undoped compound NdBa$_2$Cu$_3$O$_7$ (NBCO) revealed the in-phase $A_{1g}$ mode of the bond-buckling phonon at ~ 30 meV in accord with our data (23). RIXS experiments on NBCO and the model calculations both confirm that the EPC of the in-phase bond-buckling phonon decreases from the Brillouin zone center towards the zone boundary while the EPC of the bond-stretching phonon shows the opposite trend (20, 23-24). Comparing RIXS results between UD23 and NBCO, we found that the q-dependent EPC of the high-energy bond-stretching phonon agrees reasonably well. But an apparent discrepancy exists for the low-energy bond-buckling phonon: the dome-shaped intensity enhancement around $Q_{CDW}$ in UD23 is in stark contrast to the monotonically decreasing EPC in parent NBCO (23). The anomalous softening of the buckling phonon in YBCO was suggested to be associated with a charge-density modulation, which is now understood to be omnipresent in under-doped cuprates (22). The giant intensity anomaly of the buckling mode in UD23 may, therefore, reflect an interplay with the dispersive CDW excitations.

To elucidate the anomalous intensity enhancement of the buckling phonon, we surveyed the other parts of the phase diagram by investigating a slightly over-doped $Bi_2Sr_{1.8}La_{0.2}CuO_{6+\delta}$ superconducting compound ($T_c$ = 30 K, OD30), anticipating that the CDW correlations and EPC may be rather different than its under-doped counterpart. In Figs. 3A-3B, Cu $L_3$ and O $K$ RIXS spectra of OD30 are plotted as a function of $q_{//}$ = (H, 0). No scattering peak is observed in the quasi-elastic region at either edge across the whole accessible $q_{//}$ range. To expand on this observation, we studied RIXS under various configurations including along the (-H, 0) direction, the (H, H) direction, and in off-resonance conditions (*SI Appendix,* Fig. S9). None of them reveal any CDW signatures, making this system distinct from under-doped and the extremely over-doped Bi2201, where the CDW was found (6, 25). We illustrate these remarkable results in Figs. 3C-3D. In contrast to the well-defined CDW peak in the UD23 sample, the integrated intensity of the quasi-elastic peak in OD30 has a simple background-like profile demonstrating the complete obliteration of CDW correlations.

Correspondingly, the phonon excitations manifest differently in the absence of CDW correlations. First, the bond-stretching phonon softening is largely suppressed, regardless of whether it is probed on the Cu (Fig. 3E) or O sublattices (Fig. 3F). Second, the EPC anomaly of the bond-stretching mode, *i.e.*, the broad intensity enhancement at $q_{//} > Q_{CDW}$, diminishes in both sublattices resulting in a simple upward increase (Figs. 3G-3H). Similarly, the intensity of the bond-buckling phonon is significantly altered with no appreciable change in its dispersion (Fig. 3I). The entire dome-shaped enhancement vanishes in OD30 and a monotonically decreasing intensity profile is formed as a function of $q_{//}$ (Fig. 3J), opposite to the trend of the bond-stretching phonon.

It becomes clear now that the momentum-dependent EPC of each phonon branch in the absence of CDW correlations in OD30 coincides with that of the parent NBCO where the bond-stretching and the bond-buckling phonon intensities scale with $\sin^2(\pi H)$ and $\cos^2(\pi H)$ functions, respectively (20, 23-24). We found that these functional forms describe the observed momentum-dependent phonon intensities quite well (Figs. 3G, and 3J). The good consistency between OD30 and NBCO provides compelling evidence that the phonon intensity anomaly, in UD23, is not owing to straightforward increase in the EPC, but rather a complex reflection of the Fano interference effect induced by the dispersive CDW excitations. In Fig. 4, we illustrate a comprehensive picture of the quasi-static CDWs, the dispersive CDW excitations, and concomitant electron-phonon anomalies in the momentum-energy space. Emanating from the quasi-static CDWs, the CDW excitations quickly disperse and intersect firstly with the bond-buckling phonon at low energy. The narrow CDW excitations in the momentum space result in a dome-shaped phonon intensity enhancement closely confined around $Q_{CDW}$. After reaching higher energy and greater $q_{//}$ ($> Q_{CDW}$), the dispersive CDW excitations significantly broaden in the momentum space and intersect with the bond-stretching phonon inducing a diffused intensity anomaly. The intensity enhancement at $q_{//} > Q_{CDW}$ side is due to the momentum-dependent EPC of the bond stretching phonon. It is worth mentioning that excitations of a conventional CDW can be gapped by the periodically distorted crystal lattice or by impurities (26). Similarly, the dispersive CDW excitations in Bi2201 may exhibit a gap falling below the current detection limit.

The strong intensity anomalies of the buckling and stretching phonons in UD23 allow us to deduce the characteristic velocity of the dispersive CDW excitations. To do so, we first fitted the momentum-dependent phonon intensities and retrieved the wavevector ($Q_A$) of the maximal intensity anomaly (Fig. 3G, 3J, and *SI Appendix*, Fig. S10). By connecting $Q_A = 0.34$ *r.l.u.* with $Q_{CDW}$ = 0.259 *r.l.u.* at the Cu $L_3$ edge, we extracted the velocity of the CDW excitations, $V_{CDW\text{-stretching}}$ ~ 0.45 ± 0.05 eV Å, close to the bond-stretching phonon (~ 60 meV). Near the bond-buckling phonon (~ 30 meV), we joined the intensity anomalies at $Q_A$ = 0.238 *r.l.u.* and 0.267 *r.l.u.* with $Q_{CDW}$ = 0.25 *r.l.u.* from the O *K* edge, and obtained an averaged velocity of the CDW excitations, $V_{CDW\text{-buckling}}$ ~ 1.3 ± 0.3 eV Å. Remarkably, $V_{CDW\text{-buckling}}$ is about four times larger of $V_{CDW\text{-stretching}}$, demonstrating unambiguously that the CDW excitation disperses steeply after arising from the quasi-static CDW, then gradually flattens at higher energy. Overall, the trend extracted by the bond-stretching and the bond-buckling phonon describes funnel-shaped dispersive CDW excitations as highlighted in Fig.4. We notice that the $V_{CDW\text{-stretching}}$ in Bi2201 is comparable to that of the under-doped Bi2212 (~ 0.6 ± 0.2 eV Å) at the energy ~ 60 meV (17). Interestingly, at the energy ~30 meV, the deduced CDW excitations velocity in Bi2201 is close to that of the electron band dispersion, ~1.7 ± 0.2 eV Å, retrieved from angular-resolved photoemission data in under-doped Bi2212 (27). The similarity of the velocities may indicate comparable self-energies between the ordinary and the periodically modulated charge carriers. However, this connection should not be simply viewed as an evidence of the weak-coupling (*i.e.*, Fermi surface nesting) picture to describe the emergence of CDWs correlations. Rather, it is crucial for theoretical models to take into account the velocity values for the description of the CDWs in cuprates.

Lin *et al.* recently reported little signature of dispersive CDW excitations in LSCO compounds despite the phonon-softening across a wide doping range (19). The bond-stretching phonon intensities, with and without the presence of CDW correlation, show similar momentum dependence, in contrast to our Bi2201 RIXS data (Figs. 3G-3H, 3J). As the bond-buckling phonon intensity anomaly is much more confined around $Q_{CDW}$ differentiating drastically to the bare EPC, we corroborate the highly sensitive O $K$ RIXS in the detection of the coupling of CDW excitations with phonons and its complementary with Cu $L$ RIXS.

The co-existence of CDW at the Cu and O sublattices in Bi2201 demonstrates that the modulated charge density carries both Cu-3$d$ and O-2$p$ orbital character. REXS studies on LBCO and YBCO showed that the CDW order has $s'$- or $d$- wave symmetry which depict a bond-centered charge order naturally explaining the projection onto Cu and O sublattices (14, 28). Given the correlation length of the CDW in Bi2201 is more than an order of magnitude shorter than in LBCO (29), we suggest that the multi-orbital nature is universal to the CDWs in all hole-doped cuprates. From a theoretical perspective it is unclear whether the CDWs can be captured properly using a single-band model given the significant oxygen character of the density modulation and the phonons involved. Our work, therefore, highlights the need to use multiorbital Hubbard models (specifically 3-bands Hubbard models) to describe the cuprates in terms of electron dynamics and orders (13, 30-32).

Our findings on the rich interplay between the multi-orbital CDW and the electron-phonon anomalies in Bi2201 substantiate the existence of funnel-shaped CDW excitations dispersing in the energy-momentum space. This is a major step forward in characterising the dispersive CDW excitations comparing to the previous study on Bi2212, which was solely based on the interference effect from a single phonon branch (17). The dispersive CDW excitations are in line with the short-range dynamical charge density fluctuations found in a large portion of the phase diagram in YBCO, postulating its ubiquity in cuprates (12). Further experiments aimed at uncovering dynamical CDW excitations in different cuprates may help elucidate the relevance of CDWs for the anomalous normal state and the unconventional superconducting properties. Concerning the underlying mechanism of CDWs, the most recent CDW studies on YBCO and LSCO add mounting evidence that the Fermi-surface nesting alone is unlikely to be the primary driving force (12, 19). Moreover, the fact that the CDW dynamics interfere with the phonons points towards an important role of the EPC in the formation of the CDWs (33-36). As suggested by Refs. (20-21, 23, 37-38), the phonon intensity measured by RIXS is scaled to $M^2$, where $M$ is the EPC matrix element. A simple comparison of phonon intensities between UD23 and OD30 (Figs. 3G-3H, and 3J) would imply that $M$ at momenta far away from the phonon anomalies is comparable between two Bi2201 compounds. However, determining $M$ at $Q_{CDW}$ faces difficulties as the RIXS intensity is not anymore a simple proportion to $M^2$ but rather reflects a complex interplay with CDW excitations. It is currently challenging to extract reliably the EPC at CDW wavevector and a much more sophisticated theoretical modelling is required in future.

**Conclusion**

We combined high resolution RIXS at the O $K$- and Cu $L_3$- edges to study the quasi-static CDWs, the collective CDW excitations, and the electron-phonon coupling in Bi2201. The quasi-static CDWs are present at both Cu and O sublattices and carry comparable periodicity and correlation length implying their multi-orbital nature to be universal in hole-doped cuprates. Both the bond-stretching and the bond-buckling phonons exhibit strong anomalies concomitant to CDWs. The confined intensity anomaly of the bond-buckling phonon, together with the diffused intensity enhancement of the bond-stretching phonon, uncovered unambiguously funnel-shaped dispersive CDW excitations. The significant interference effects suggest that CDWs are intimately connected to the electron-phonon coupling, which needs to be considered as a crucial element for the underlying mechanism leading to CDWs in cuprates.

## Materials and Methods

**Sample growth and characterization.** High-quality single crystals of UD23 and OD30 $Bi_2Sr_{2-x}La_xCuO_{6+\delta}$ cuprates with $x = 0.6$ and $0.2$, respectively, were grown by the traveling-solvent floating-zone method. The as-grown samples were annealed at 650°C in oxygen atmosphere for two days to improve sample homogeneity. The samples were pre-characterized and aligned using Laue diffraction prior to RIXS experiments (*SI Appendix*, Fig. S1). Superconducting transition temperature $T_c$ are 23 K and 30 K for UD23 and OD30 (*SI Appendix*, Fig. S1) which are consistent to a doping concentration of $p \sim 0.13$ and $\sim 0.18$, respectively (39).

**High resolution RIXS measurements on UD23 and OD30.** High-resolution RIXS experiments were performed at the I21-RIXS beamline at Diamond Light Source, United Kingdom. Samples were cleaved in air prior the transfer into the sample load lock vacuum chamber. The experimental geometry is sketched in *SI Appendix*, Fig. S2. All samples were aligned with the surface normal (001) lying in the scattering plane. X-ray absorption was measured using the total electron yield (TEY) method by recording the drain current from the samples. For RIXS measurements, linear σ and π polarized X-rays were used. The total energy and momentum resolution were about 40 meV (26 meV) (FWHM) and ±0.012 Å$^{-1}$ (±0.006 Å$^{-1}$) at the Cu $L_3$ (O $K$) edge, respectively. To enhance the RIXS throughput, a special paraboloidal mirror is implemented in the main vacuum chamber. The RIXS spectrometer was positioned at a fixed scattering angle of 154° resulting in a maximal total momentum transfer value Q, of ~ 0.92 Å$^{-1}$ (0.52 Å$^{-1}$) at the Cu $L_3$ (O $K$) edge. The projection of the momentum transfer, $q_{//}$, in the *a-b* plane was obtained through varying the grazing incident angle of the sample owing to the quasi two-dimensional CDW in Bi2201 system. We use the pseudo-tetragonal unit cell with $a = b = 3.86$ Å and $c = 24.69$ Å for the reciprocal space mapping. The momentum transfer **Q** is defined in reciprocal lattice units (r.l.u.) as $\mathbf{Q} = H\mathbf{a}^* + K\mathbf{b}^* + L\mathbf{c}^*$ where $\mathbf{a}^* = 2\pi/a$, $\mathbf{b}^* = 2\pi/b$, and $\mathbf{c}^* = 2\pi/c$. All measurements were done at 20 K under a vacuum pressure of about 5×10$^{-10}$ mbar. To confirm the observation of CDW, the RIXS measurements were repeated on a second sample for UD23 with the same experimental set-up. Detailed results are summarized in *SI Appendix*, Fig. S4.

**Data analysis and data fitting.** All RIXS spectra have been normalized by the counting time and corrected for self-absorption effects through the procedure described in *SI Appendix*. The zero-energy positions of RIXS spectra were determined by comparing to reference spectra recorded from the amorphous carbon tapes next to the sample for each $q_{//}$ position. They were finely adjusted through the Gaussian fitting of each elastic peak. The intensity of CDW as a function of $q_{//}$ results from the integration of spectra within an energy window of ± 30 meV and was fitted with a Gaussian profile and a power-law function as a background. To quantify the phonon excitations, the fitting model for the Cu $L_3$ spectra (from -100 meV up to 800 meV) consists of a Gaussian (elastic peak) with a width constrained to the instrumental energy resolution, a Lorentzian (bond-stretching phonon), a damped harmonic oscillator model to account for the paramagnon at high energy and a linear background. For the O $K$ RIXS data where the elastic peaks are generally much stronger, an instrumental energy resolution limited Gaussian function is firstly used to fit the spectra ranging from -100 meV to 25 meV, then two Gaussians (for the buckling and bond-stretching phonon modes) and a linear trend (for the background) are applied to fit the residual spectra from -100 meV to 150 meV after removing the fitted elastic peaks. We note that the fitted phonon parameters in the O $K$ spectra are generally consistent with a global fitting containing three Gaussian functions, except the central position of the bond-buckling mode is ~ 10 meV lower than the fitted central position of ~ 30 meV presented in the main text. Error bars of the phonon dispersions are determined by a combination of the uncertainty of determining the zero-energy position and standard deviations from the fit. For the integrated RIXS intensities of CDW and phonons, error bars are based on the noise level of measured spectra.

**Acknowledgments**

We thank W.-S. Lee for fruitful discussions. J. Li acknowledges Diamond Light Source (United Kingdom) and the Institute of Physics in Chinese Academy of Sciences (China) for providing funding Grant 112111KYSB20170059 for the joint Doctoral Training under the contract STU0171. H. C. Robarts acknowledges funding and support from the Engineering and Physical Sciences Research Council (EPSRC) Centre for Doctoral Training in Condensed Matter Physics (CDT-CMP), Grant No. EP/L015544/01 and Grant EP/R0111141/1. H.D. acknowledges the financial support from the National Natural Science Foundation of China (Grant 11888101), and the Ministry of Science and Technology of China (Grant 2016YFA0401000). S. J. acknowledges support from the National Science Foundation under Grant No. DMR-1842056. J. P. acknowledges financial support by the Swiss National Science Foundation Early Postdoc Mobility fellowship project number P2FRP2_171824 and PostDoc Mobility project number P400P2_180744. All data were taken at the I21 RIXS beamline of Diamond Light Source (United Kingdom) using the RIXS spectrometer designed, built and owned by Diamond Light Source. We acknowledge Diamond Light Source for providing the beamtime on Beamline I21 under proposals NR19886, NR21184 and NT21277. We acknowledge Thomas Rice for the technical support throughout the beamtime. We thank G. B. G. Stenning and D. W. Nye for help on the Laue instrument in the Materials Characterization Laboratory at the ISIS Neutron and Muon Source.

**Competing interest statement:**

The authors declare no competing interests.

**Author Contributions**

K.-J.Z. conceived the project and led the experiments. J.L., A.N., K.-J.Z., J.P., H.C.R., A.C.W., and M.G.F. performed measurements at Diamond Light Source. J.L. and K.-J.Z. analysed the data with support from A.N., J.P., H.D., R.C., and S.J.. H.E. and D.S. synthesized and prepared samples for the experiments. J.L. performed offline sample characterization. K.-J.Z. and J.L. wrote the manuscript with input from all the authors.

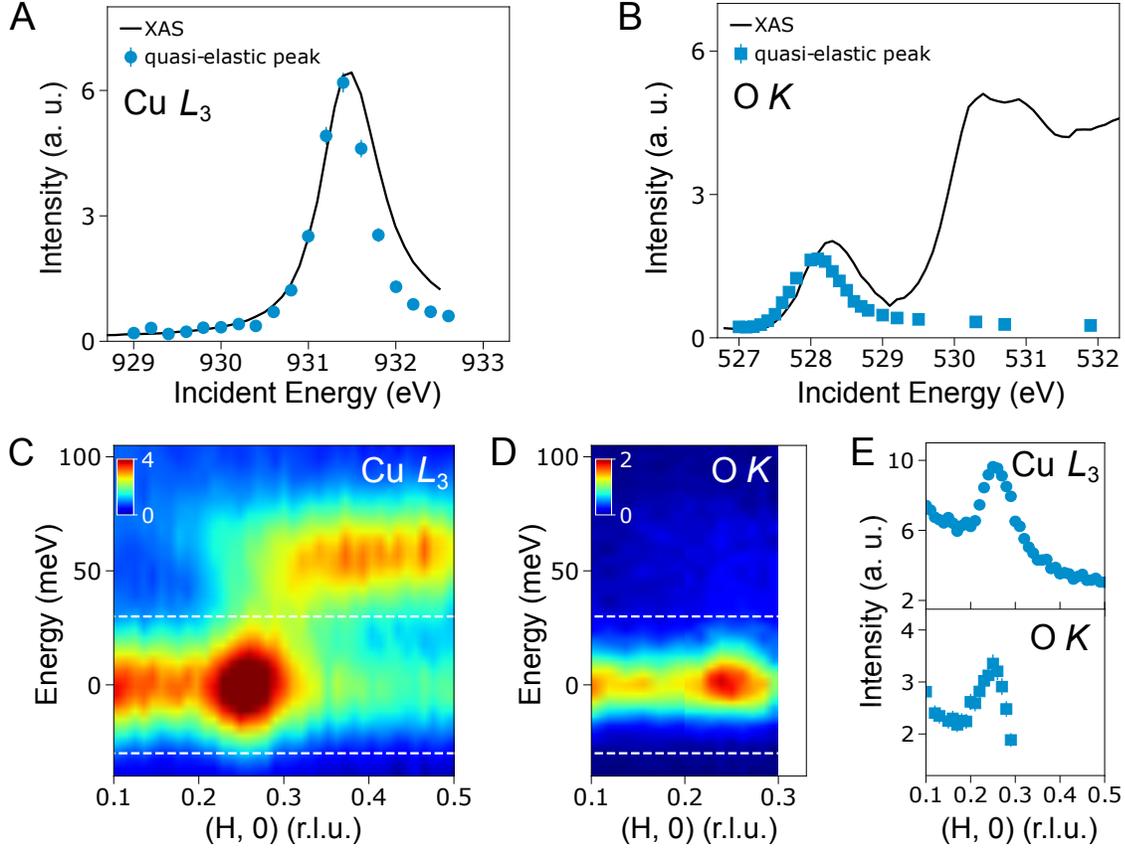

**Figure 1. CDW in under-doped UD23. A** and **B**, Solid black lines: Cu $L_3$- and O $K$- edges XAS of UD23 sample collected with σ incoming polarization in normal incidence geometry (see the experimental setup in Fig. S2). Blue circles (squares): Energy-dependent scattering intensities of the quasi-elastic peaks at, the Cu $L_3$-edge (O $K$-edge) at $q_{//} = 0.26$ *r.l.u.* (0.25 *r.l.u.*). **C** and **D**, RIXS intensity maps excited at, the Cu $L_3$ resonance of 931.6 eV and the O $K$-edge hole-peak resonance of 528.4 eV as a function of energy and $q_{//}$ along the (H, 0) direction. **E**, Integrated intensity of the quasi-elastic peak, at the Cu $L_3$- and O $K$- edges, within an energy window (± 30 meV) marked by the two white dashed lines in **C** and **D**.

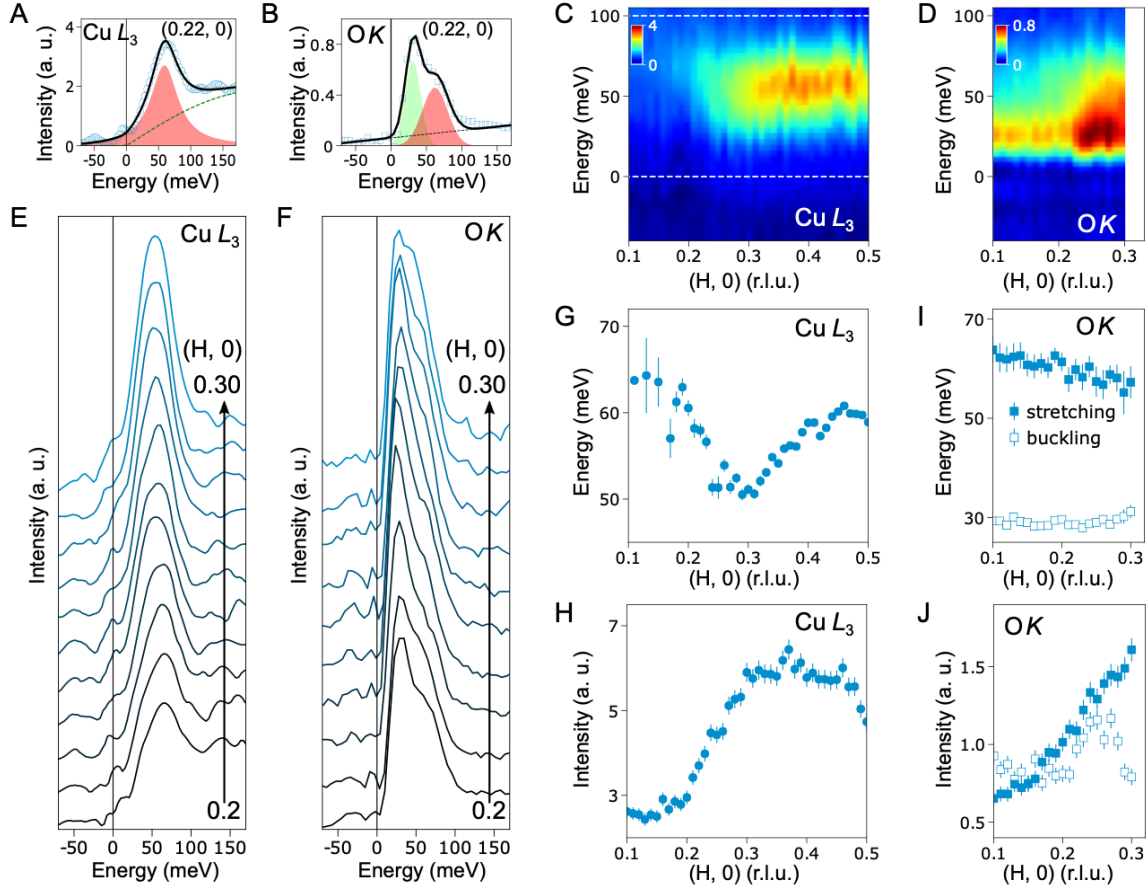

**Figure 2. Phonons in under-doped UD23. A** and **B**, Fitting examples of the inelastic excitation at the Cu $L_3$- and O $K$- edges. Fitting details are described in the Methods section. **C** and **D**, Cu $L_3$ and O $K$ RIXS intensity maps as shown in Fig. 1**C** and 1**D** with the fitted elastic peak subtracted. **E** and **F**, Cu $L_3$ and O $K$ RIXS spectra after removing the fitted elastic peaks at selected momentum transfer values ranging from $q_{//} = 0.2$ *r.l.u.* to 0.3 *r.l.u.* along the (H, 0) direction. **G** and **H**, The fitted dispersion and the integrated intensity of the bond-stretching phonon excitations at the Cu $L_3$-edge as a function of momentum. The phonon spectral weight is integrated within the energy window (0-100 meV) illustrated by the white dotted line in **C**. **I** and **J**, The dispersion and integrated intensity of the bond-stretching phonon (the high-energy peak, filled markers) and the bond-buckling phonon (the low-energy peak, open markers) at the O $K$-edge. The dispersion is extracted from the fitting. Phonon intensity is the area of the curve fitting each phonon mode.

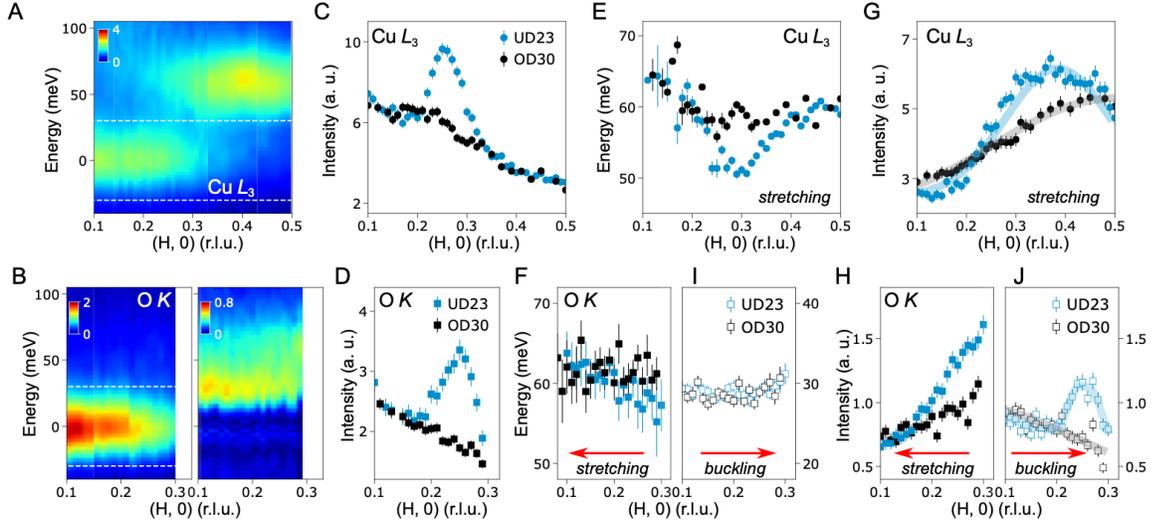

**Figure 3. CDW and phonons studies in over-doped OD30. A** and **B,** Cu $L_3$ and O $K$ RIXS intensity map as a function of energy and $q_{//}$ along the (H, 0) direction, respectively. The map on the right-hand side of **B** is the residual excitations after subtraction of the fitted elastic peaks. **C** and **D**, Integrated intensities of the quasi-elastic peaks at the Cu $L_3$- (black dots) and O $K$- (black squares) within the energy window (± 30 meV) defined by the two white dashed lines in **A** and **B**. Integrated elastic peaks intensities in UD23 from the same energy range are shown in blue for comparison. **E**, Dispersion of the bond-stretching phonon excitations at the Cu $L_3$, in OD30 (black dots) extracted from the fitting. **F** and **I**, Dispersion of the bond-stretching phonon (black filled squares) and bond-buckling phonon (black open squares) excitations at the O $K$-edge. For **E**, **F**, and **I**, Phonon peak positions extracted from UD23 sample are displayed in blue for comparison. **G**, Integrated intensity of the bond-stretching phonon at the Cu $L_3$-edge within the same energy window (0-100 meV) as for UD23 sample after the removal of the elastic peak. **H** and **J**, Bond-stretching and bond-buckling phonon intensities through the integration of the area of each fitted phonon mode at O $K$-edge. For **G**, **H** and **J**, phonon intensities in UD23 are displayed in blue for comparison. Fits of the phonon intensities are superimposed on top of experimental data in **G** and **J**.

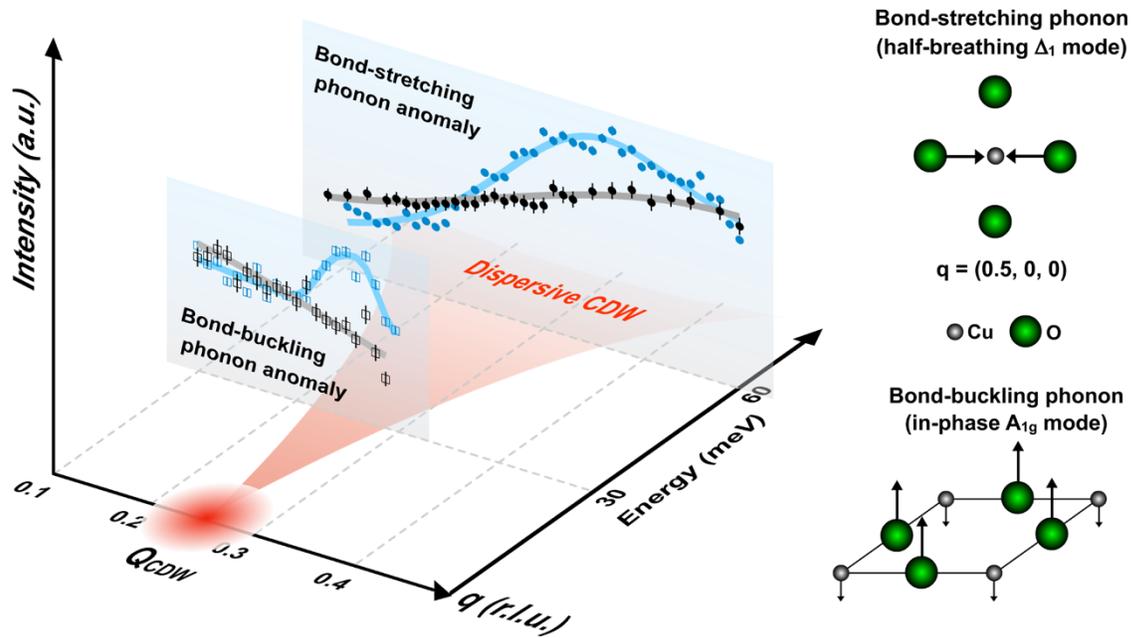

**Figure 4. Quasi-static CDWs, their dispersive excitations, and concomitant electron-phonon anomalies.** A sketch of the quasi-static CDW, the dispersive CDW excitations, and their interplay with the bond-stretching and the bond-buckling phonon modes. Blue and grey markers are the experimental phonon intensities of UD23 and OD30 samples, respectively, as shown in Fig. 3G and 3J. Solid lines are fitting results. Sketches on the right illustrate the bond-stretching phonon (the half-breathing $\Delta_1$ mode) at q = (0.5, 0, 0) and the bond-buckling phonon (in-phase) $A_{1g}$ mode.

Supplementary Information for

# Multi-orbital charge density wave excitations and concomitant phonon anomalies in Bi$_2$Sr$_2$LaCuO$_{6+\delta}$

Jiemin Li, Abhishek Nag, Jonathan Pelliciari, Hannah Robarts, Andrew Walters, Mirian Garcia-Fernandez, Hiroshi Eisaki, Dongjoon Song, Hong Ding, Steven Johnston, Riccardo Comin, Ke-Jin Zhou[*]

**Correspondence to: kejin.zhou@diamond.ac.uk**

This PDF file includes:

    Supplementary Text

    Supplementary Figures

    Supplementary Tables

    Supplementary References

**Sample Information**

we focused on Bi2201 superconductors with two different doping levels: (1) $Bi_2Sr_{1.6}La_{0.4}CuO_{6+\delta}$ (UD23), p ~ 0.13, Tc ~ 23 K; (2) $Bi_2Sr_{1.8}La_{0.2}CuO_{6+\delta}$ (OD30), p ~ 0.18, Tc ~ 30 K, as indicated by the black arrows shown in Fig. S1D. The measured Laue patterns of the samples are displayed in Fig.S1A and Fig. S1B. The superconducting transition temperature, Tc, were determined from the magnetization measurements shown in Fig. S1C and S1D. Please note that 10 Oe is already a low field limit therefore the use of the magnetic field values (10 and 1 Oe) does not make considerable difference for determining Tc between two samples. A set of super-structural (SS) points due to the structural distortion of BiO planes are clearly resolved and used as the reference to orient samples.

**RIXS Scattering Geometry**

Fig. S2A shows a sketch of the RIXS experimental scattering geometry. Samples were mounted such that the *a*-axis and *c*-axis lay in the horizontal scattering plane while the *b* axis was perpendicular to the scattering plane. During the measurements, the spectrometer was fixed to $\Omega$ = 154° and the projection of the momentum transfer was obtained through varying θ angle around the sample b axis. Both σ and π polarized incoming X-rays were employed to probe samples with no polarization analysis for the outgoing scattered X-rays. To get a clean sample surface for X-ray measurements, we cleaved the samples in air prior transferring into the load lock vacuum chamber. Throughout the whole experiments, the reciprocal space is defined with a unit cell of *a* = *b* = 3.86 Å and *c* = 24.69 Å. We display all the data as reciprocal lattice units. Accessible momentum space in the first Brillouin Zone at the Cu $L_3$-edge (green shaded circle) and the O *K*-edge (magenta shaded circle) are shown in Fig. S2B. The measurements were done mainly along the (H, 0), (0, H) and (H, H) directions.

**The self-absorption correction for RIXS spectra**

The self-absorption can distort the spectral line profile severely. To eliminate this effect, we follow the procedure described in Ref. S1. Considering that the elastic peak and the low-energy phonon do not involve the spin-flip process meaning the X-ray polarization is preserved, *i.e.*, $I_{\sigma \to \pi} = I_{\pi \to \sigma} = 0$, the intensity correction can be described as:

$$I_{\sigma}^{corr} = I_{\sigma \to \sigma}^{corr} = I_{\sigma}/C_{\sigma \to \sigma} = I_{\sigma} \times \left( \mu_{i\sigma} + \mu_{f\sigma} \times \frac{-\vec{k}_i \cdot \vec{n}}{\vec{k}_f \cdot \vec{n}} \right) = I_{\sigma} \times f_a \times \left( 1 + \frac{\sin\theta}{\sin(\Omega - \theta)} \right),$$

$$I_{\pi}^{corr} = I_{\pi \to \pi}^{corr} = I_{\pi}/C_{\pi \to \pi} = I_{\pi} \times \left( \mu_{i\pi} + \mu_{f\pi} \times \frac{-\vec{k}_i \cdot \vec{n}}{\vec{k}_f \cdot \vec{n}} \right)$$

$$= I_{\pi}$$

$$\times \left( (f_a \sin^2\theta + f_c \cos^2\theta) + (f_a \sin^2(\Omega - \theta) + f_c \cos^2(\Omega - \theta)) \times \frac{\sin\theta}{\sin(\Omega - \theta)} \right)$$

Here, $\mu_i$ ($\mu_f$) and $\vec{k}_i$ ($\vec{k}_f$) characterize the absorption coefficient and the direction of the incoming (outgoing) X-rays, respectively. $\vec{n}$ is the surface normal direction, $f_a$ and $f_c$ are the Cu or O atom scattering factor along *a* or *c*-axes which can be extracted from the XAS with $f_a \gg f_c$ due to the quasi-two-dimensional nature of the sample. It is clear that the self-absorption correction for σ polarized RIXS spectra is only related to the scattering geometry and scaled by the in-plane scattering factor $f_a$. Whereas for the π polarized spectra, both $f_a$ and $f_c$ contribute to the correction. The self-absorption correction was applied to all RIXS data presented in the paper.

**CDW excitations in various configurations**

We present the observation of the CDW excitation in various experimental configurations alongside the results shown in the main text. In Fig. S3 we show the RIXS map excited using the π polarized incident X-rays along the (H, 0) direction, RIXS map excited using the σ polarized incident X-rays along the (0, H) direction, and RIXS map excited using the σ polarized incident X-rays along the (H, H) direction. We see clear CDW scattering peak along the (H, 0) and (0, H) directions regardless the polarization of the incident X-rays. However, no CDW scattering peak is formed along the (H, H) direction consistent to the RXS study on Bi2201 compounds [S2]. Integrated intensity within the quasi-elastic region (± 30 meV) defined by the white dashed lines was summarized in Fig. S3d. CDW peak along the (H, 0) or the (0, H) direction has a characteristic wavevector at $q_{//}$ ~ 0.26 r.l.u..

**Repeatability of the presence of the CDW peak in UD23**

To confirm the observation of CDW at both the Cu $L_3$- and O $K$- edges in UD23, we repeated the measurements on a second sample of UD23 which show again CDW scattering peak at both edges. Using the fitting analysis described in the Methods of the main text, we show the fitting results in the table below and the integrated intensity of the quasi-elastic peak (± 30 meV) in Fig. S4. The error bars shown in the table are the convolution of the instrumental momentum resolution and the standard deviation of the fitting errors.

**Fittings of the bond-stretching and the bond-buckling phonons**

We show detailed fitting results of the bond-stretching and the bond-buckling phonons in $q_{//}$ dependent spectra in UD23 at the Cu $L_3$ and the O $K$ in Fig. S5, and Fig. S6, respectively. For the Cu $L_3$-edge data, a Gaussian function with the instrumental resolution is used for fitting the elastic peak, a Lorentzian function is applied to the bond-stretching phonon, the tail of paramagnon is fitted by a damped harmonic oscillator multiplied by the Bose factor, and a linear function is used for fitting the general background. For the O $K$-edge data, two Gaussian functions are used to fit the residual spectra after subtracting the fitted elastic peak to track the bond-stretching and the bond-buckling phonon modes.

**Ultra-high energy resolution O *K*-edge RIXS of UD23**

The fitting analysis for the O *K* RIXS spectra of UD23 shows that the phonon excitations contain a two-peak structure though the lower-energy phonon is less visible comparing to the higher-energy phonon peak in the raw data. To confirm the observation, we re-measured the same UD23 sample using a higher energy resolution (FWHM = 18 meV) set-up than the normal energy resolution (FWHM = 26 meV) set-up used for the main measurements. Fig. S7 shows the comparison of the raw RIXS spectra at $q_{//}$ of (0.23, 0). A low energy phonon excitation below 50 meV is clearly resolved in the higher energy resolution RIXS spectrum.

**Extracted dispersion of the bond-stretching phonon in UD23 at Cu $L_3$**

Using the same fitting analysis described in the Methods of the main text, we fitted the excitations at the Cu $L_3$-edge of UD23 measured along the (H, 0) direction using π polarized X-rays and along the (H, H) using σ polarized X-rays. Extracted dispersion of the bond-stretching phonon along the (H, 0) shows quite good consistency between σ and π polarizations. In particular, both develop a softening near $Q_{CDW}$ despite totally different $q_{//}$-dependent phonon intensity distributions (Fig. S3A vs Fig. 1C). The extracted phonon dispersion along the (H, H) direction shows no softening with slightly a higher excitation energy comparing to that along the (H, 0) direction. Results are shown Fig. S8.

**Detailed examination of the quasi-elastic peak in OD30 at Cu $L_3$- and O *K*- edges**

At the resonance of the Cu $L_3$-edge (931.6 eV) and the hole peak (528.4 eV) of O *K*-edge in OD30 sample, no CDW is observed as shown in Fig. 3. To further explore the existence of the CDW, we performed the Cu $L_3$ RIXS measurements at the resonance (931.6 eV) along the (-H, 0) (Fig. S9A), at an energy off-resonance (932 eV) along the (H, 0) direction (Fig. S9B), and at the resonance (931.6 eV) along the (H, H) direction (Fig. S9C). For O $K$ RIXS, we collected data at an energy off-resonance of the hole peak (528.1 eV) (Fig. S9E). Figures S9D and Fig. S9F summarize the integrated intensity of the quasi-elastic peak within the energy window (± 30 meV) defined by white dashed lines at the Cu $L_3$ and O $K$ edges, respectively. None of them shows any signature of CDW scattering peak. Strong CDW peak from UD23 is shown in both Fig. S9D and Fig. S9F for comparison.

**Fittings of the momentum-dependent intensities of the bond-stretching and the bond-buckling phonons in UD23**

To extract the velocity of dispersive CDW excitations in UD23, we fitted the momentum-dependent intensities of the bond-stretching (bond-buckling) phonon mode obtained from the Cu $L_3$ (O $K$) RIXS. We know that the momentum-dependent intensities of the bond -stretching and -buckling phonons in OD30 follow $\sin^2(\pi H)$ and $\cos^2(\pi H)$ functions, respectively. For the bond-stretching phonon in UD23, we use $A\sin^2(\pi H) + C$, where A and C are constants, to describe the background signal and a Gaussian to account for the anomalous phonon profile. Likewise, for the bond-buckling phonon, $A'\cos^2(\pi H) + C'$ function is used to fit the background and two Gaussians are used to fit two phonon anomalies. We obtained the bond-stretching phonon anomaly at $Q_A$ = 0.34 ± 0.007 *r.l.u.*. For the bond-buckling phonon, the two anomalies are at $Q_A$ = 0.238 ± 0.004 *r.l.u.*, and $Q_A$ = 0.267 ± 0.004 *r.l.u.*. The velocity of dispersive CDW excitations near the bond-buckling phonon is an averaged value based on two phonon anomalies. Fittings are shown in Fig. S10.

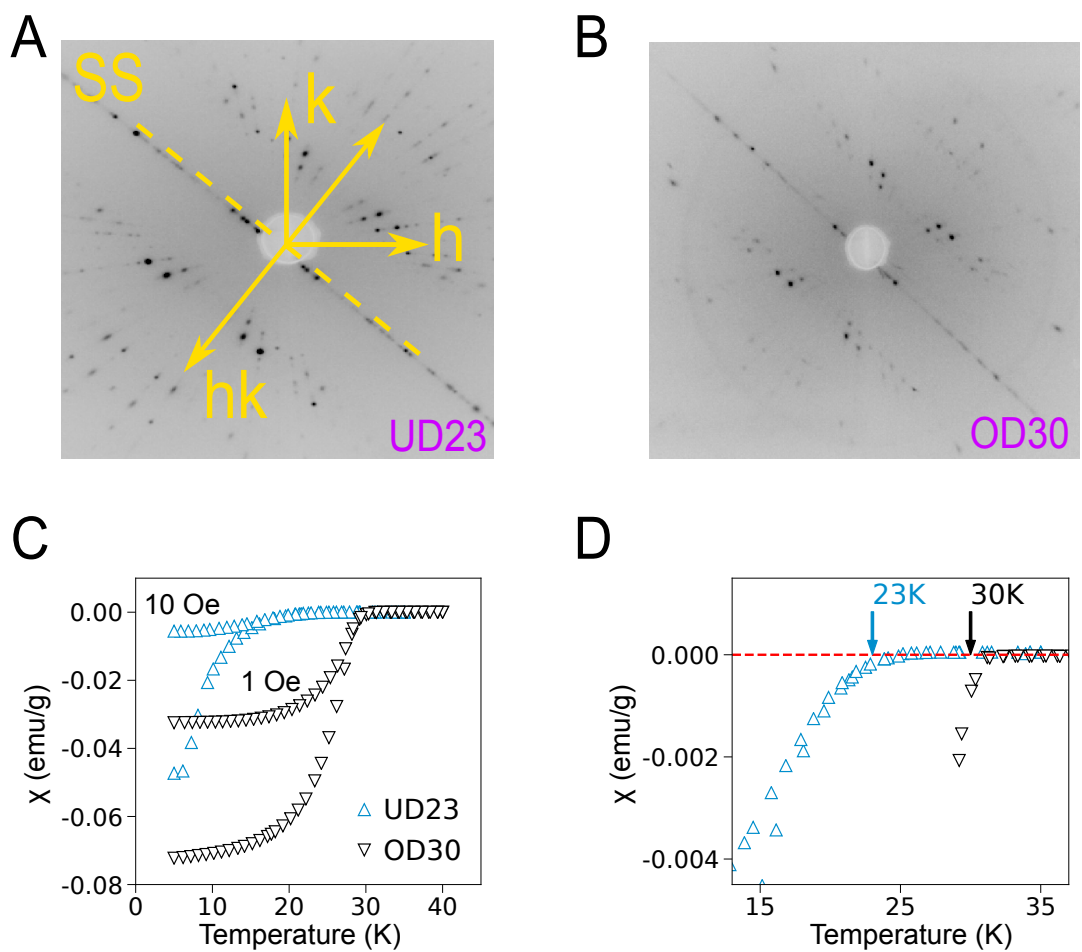

**Figure S1. Sample Information**. **A** and **B**, The Laue patterns of UD23 and OD30 samples, respectively. The *h* and *hk* denotes (H, 0) and (H, H) direction, respectively. The super-structure (SS) diffraction points of BiO structural distortion is perpendicular to the *hk* direction. **C** and **D**, Magnetization results for UD23 and OD30 samples.

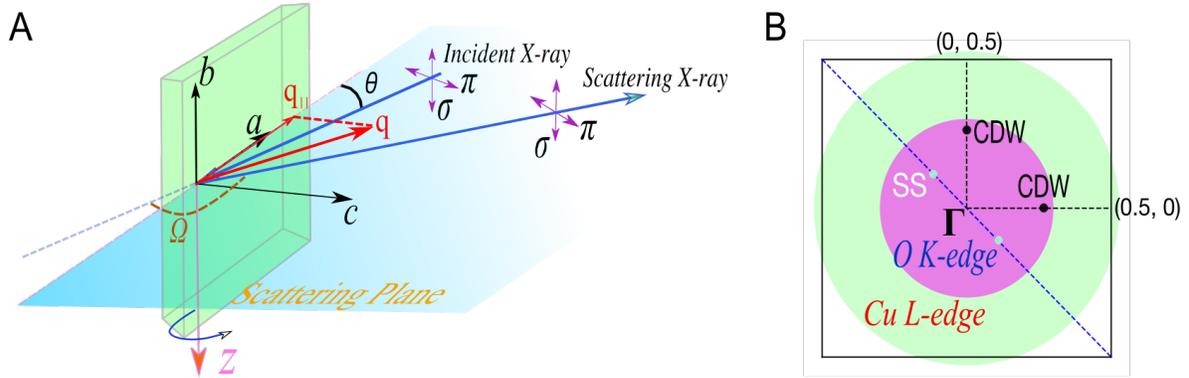

**Figure S2. RIXS experimental set-up. A**, Scattering geometry. Blue arrows define the incident and scattered X-rays. Purple cross illustrates the linear polarization of the incident X-rays. The red arrow, q, denotes the total photon momentum transfer. Its projection onto the samples a-axis is described by $q_{//}$. $\Omega$ defines the two-theta angle between the incident and scattered X-ray beam. a, b, and c define the primary sample lattice axes. The in-plane $q_{//}$ projection along the negative direction, *i.e.*, (-H, 0), and the positive direction, *i.e.*, (H, 0), is defined when $\theta<\Omega/2$, and $\theta>\Omega/2$, respectively. **B**, Accessible momentum space at the Cu $L_3$-edge (green) and the O $K$-edge (magenta) in the first Brillouin Zone. The SS direction is highlighted by a dashed blue line and the CDW wavevectors are highlighted by black dots.

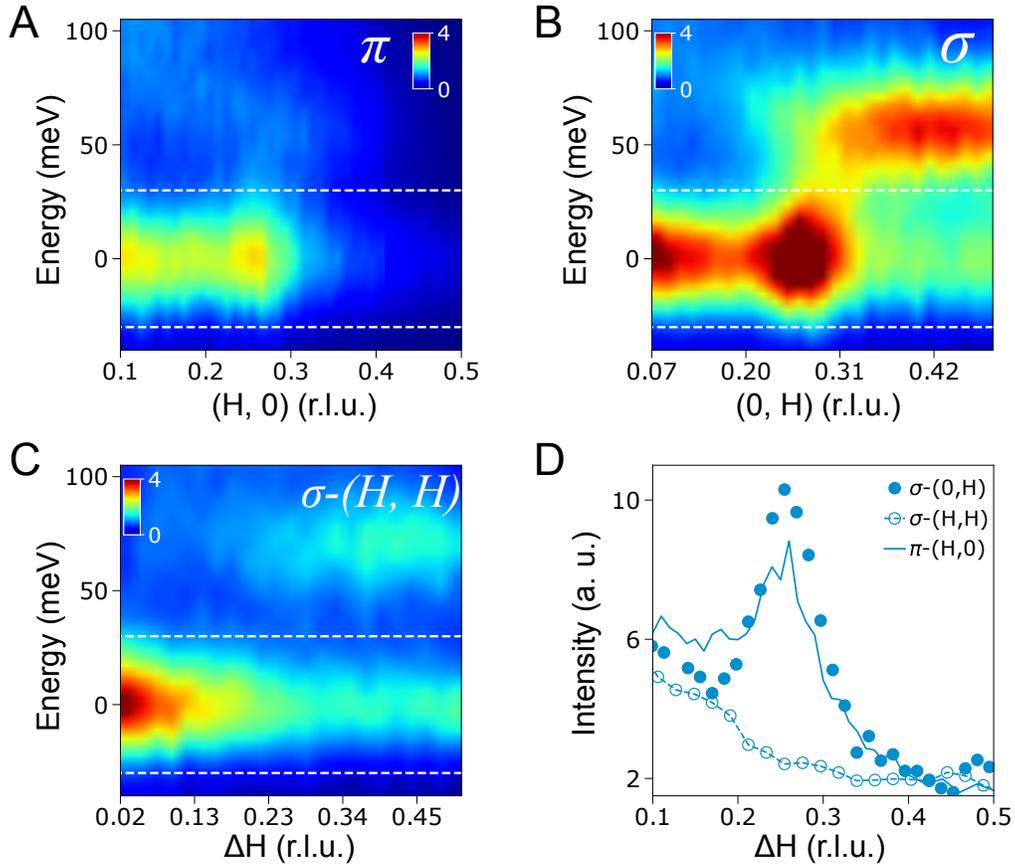

**Figure S3. RIXS intensity maps from various configurations. A**, Data obtained using $\pi$ polarized incident X-rays. **B** and **C**, RIXS intensity map using the $\sigma$ polarized incident X-rays along $(0, H)$ and $(H, H)$ directions, respectively. **D**, Integrated intensities within the white dashed line defined in **A**, **B** and **C**, as a function of the momentum transfer. Note that the horizontal axes in **C** and **D** are labeled by the absolute value of the in-plane momentum transfer, $\Delta H = \sqrt{H^2 + H^2}$.

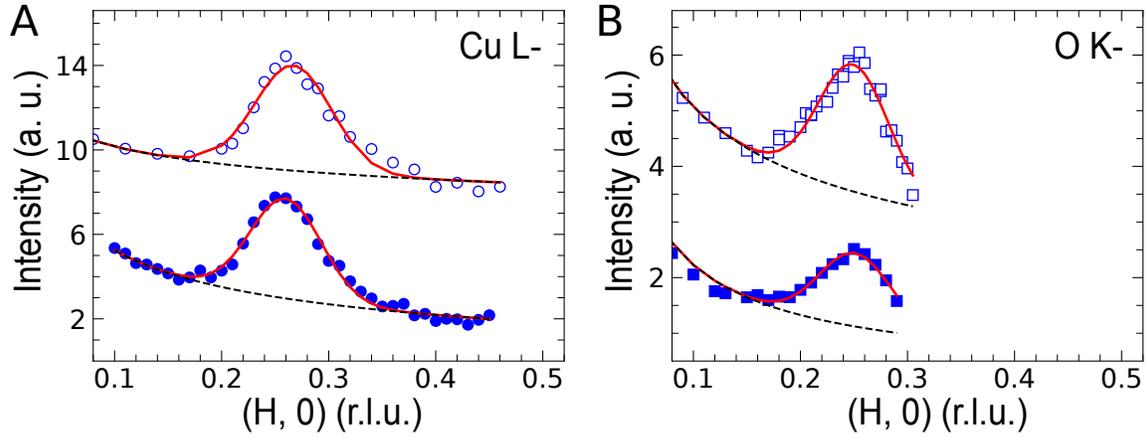

**Figure S4. Integrated intensity of the quasi-elastic region of CDW in UD23.** The red line is the fitting from a Gaussian and a power-law function (black dashed line). Filled markers are for the Sample 1 shown in the main text, while open markers stand for the Sample 2. Data are shifted vertically for comparison.

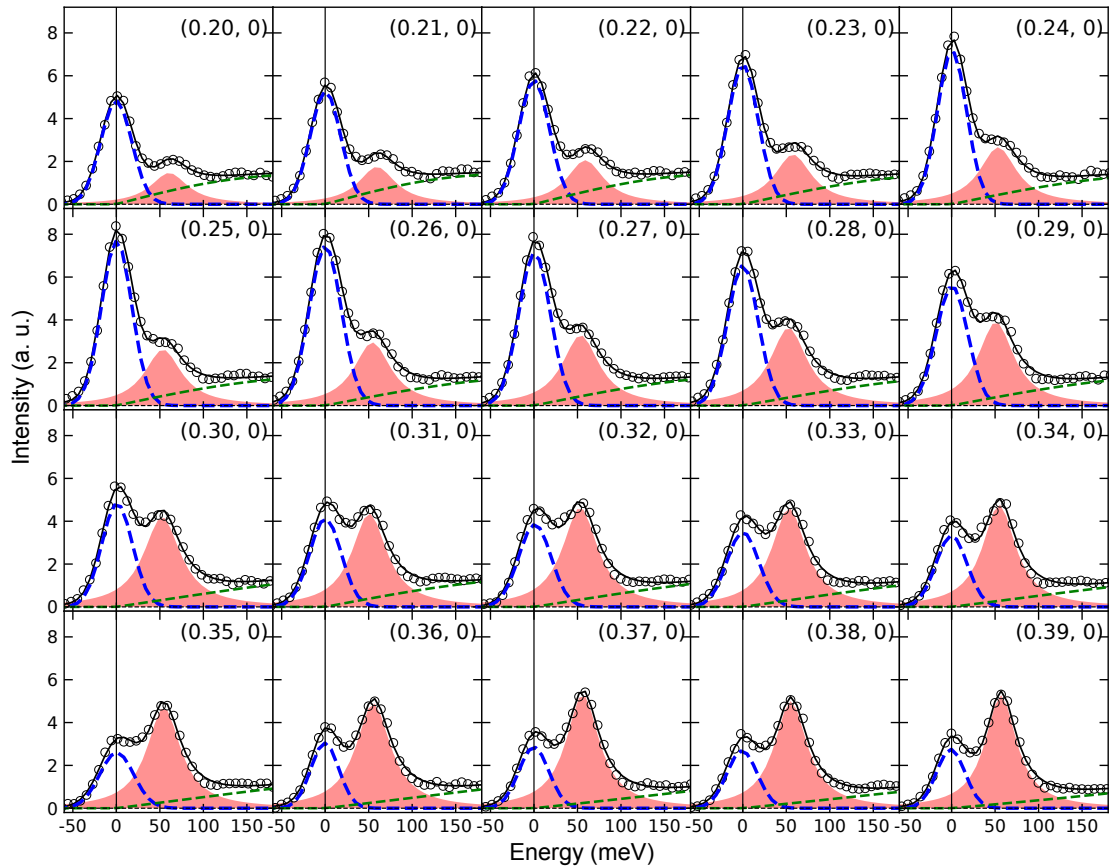

**Figure S5. Fittings of the bond-stretching phonon at the Cu $L_3$-edge in UD23.** Blue dashed line is the fit to the elastic peak, the red shaded area represents the fit of the bond-stretching phonon, the green dashed line is the fit to the tail of paramagnon excitations, and the black dashed line stands for the fit of the background.

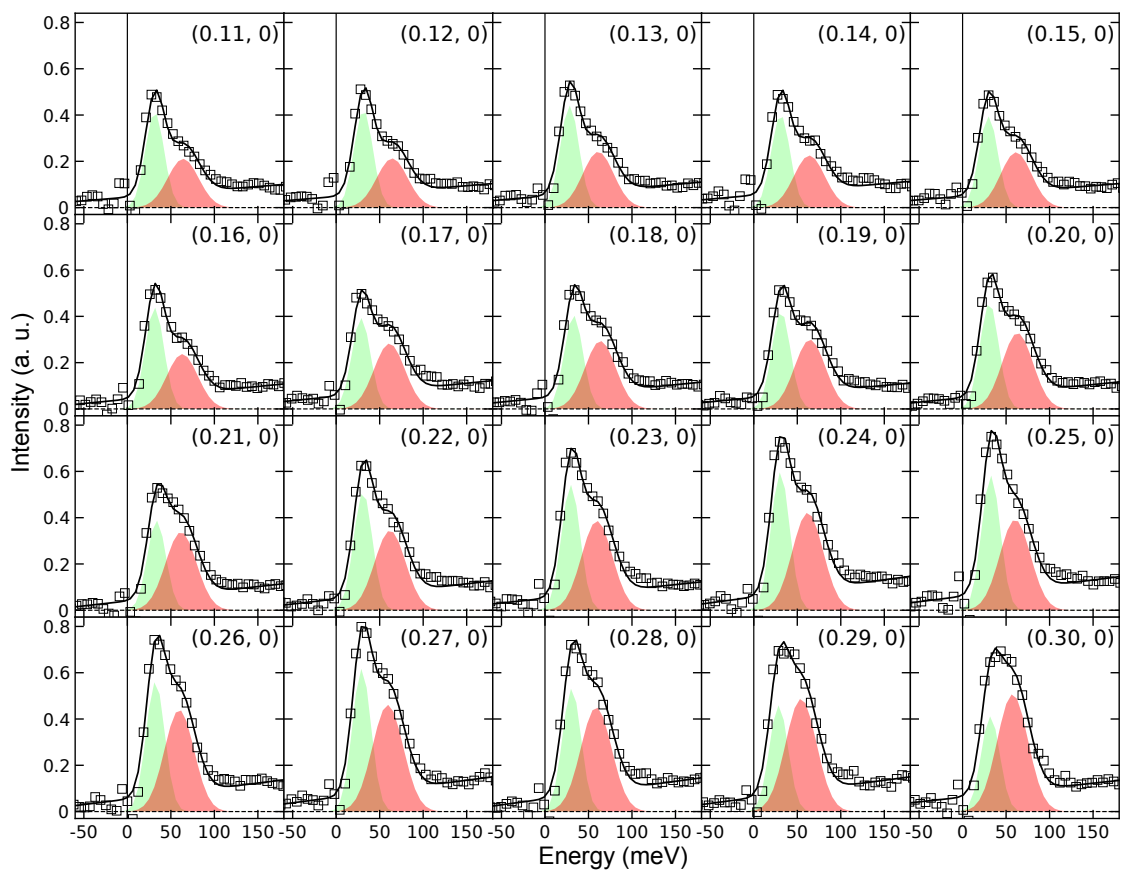

**Figure S6. Fitting of the bond-stretching and the bond-buckling phonons at the O *K*-edge in UD23.** Green shaded peak represents the fit of the bond-buckling phonon, the red shaded peak represents the fit of the bond-stretching phonon.

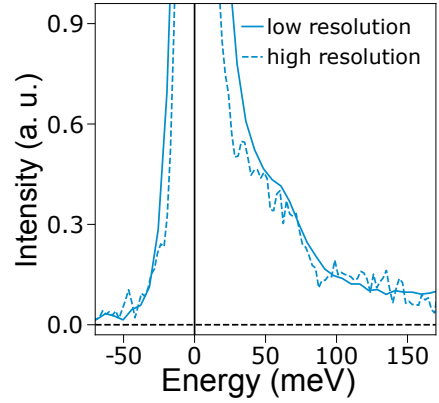

**Figure S7. O *K* RIXS data of UD23 sample collected using two energy resolutions.** The solid and dashed line represents data collected using an energy resolution (FWHM) of 26 meV and 18 meV, respectively. Both spectra were collected at a fixed momentum transfer $q_{//}$ = (0.23, 0).

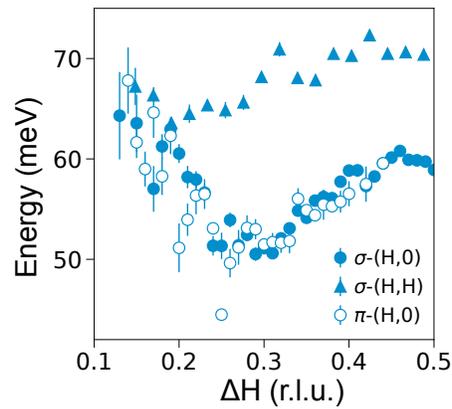

**Figure S8. Phonon dispersions of UD23.** Extracted phonon dispersion of UD23 sample1 measured along various high symmetry directions in the reciprocal space.

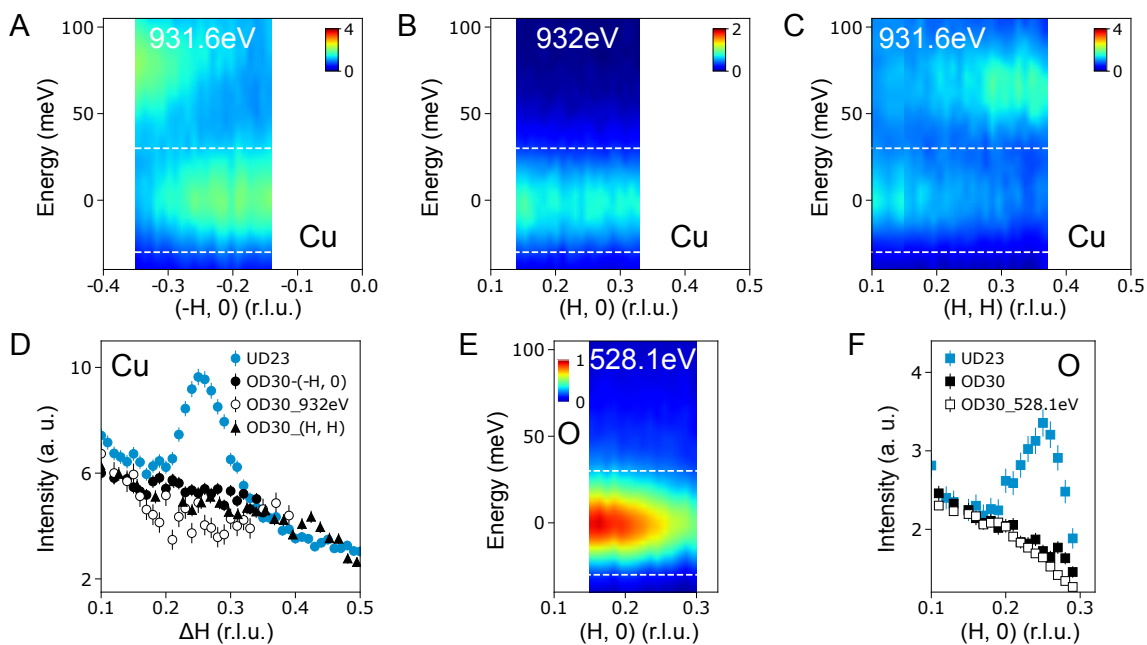

**Figure S9. Examination of CDW in OD30 from different experimental configurations. A** is along (-H, 0) direction at the resonant photon energy of 931.6 eV. **B** is along the (H, 0) direction at an off-resonant energy of 932 eV. **C** shows the result along the (H, H) direction at the resonant energy of 931.6 eV. **E** shows the off-resonance (528.1 eV) RIXS map at O *K*-edge. **D** and **F,** comparison of integrated quasi-elastic peaks of various configurations at the Cu $L_3$- and O *K*-edges, respectively. CDW peak of UD23 sample is shown in **D** and **F**.

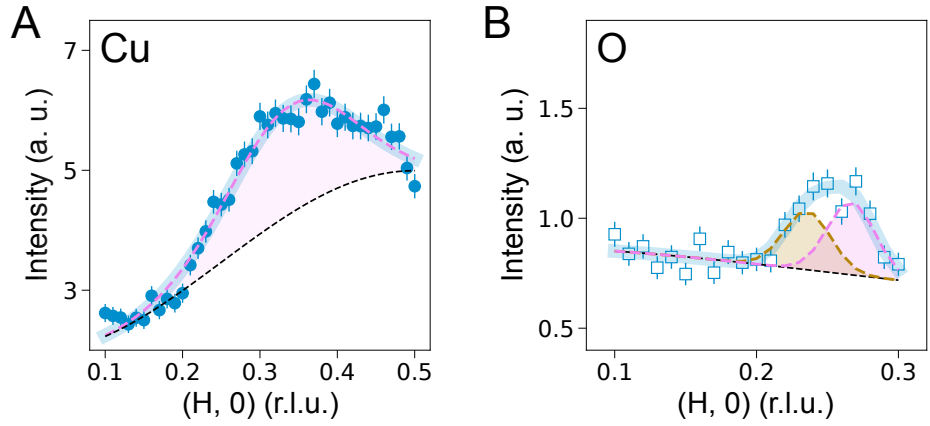

**Figure S10. Fitting of momentum-dependent phonon anomalies in UD23. A** is the fitting of the bond-stretching phonon mode. The pink dashed line is the fitted Gaussian peak and the black dashed line represents the background. **B** is fitting of the bond-buckling phonon mode. The pink and brown lines are two Gaussian peaks and the black dashed line stands for the background.

**Table S1.** Repeatability of CDW peak in UD23 samples.

|  | Cu $L_3$-edge | | O $K$-edge | |
|---|---|---|---|---|
|  | **$Q_{CDW}$ (*r.l.u.*)** | **FWHM (*r.l.u.*)** | **$Q_{CDW}$ (*r.l.u.*)** | **FWHM (*r.l.u.*)** |
| Sample1 | 0.259 ± 0.006 | 0.078 ± 0.006 | 0.25 ± 0.003 | 0.076 ± 0.006 |
| Sample2 | 0.266 ± 0.006 | 0.084 ± 0.007 | 0.25 ± 0.003 | 0.078 ± 0.006 |